\documentclass[
 reprint,
superscriptaddress,
 amsmath,amssymb,
 aps,
pra,
]{revtex4-1}
\usepackage{color}
\usepackage{amsfonts}
\usepackage{tikz-cd}
\usepackage{graphicx}
\usepackage{amsmath,amssymb,graphicx}
\usepackage{caption}
\usepackage{subcaption}
\usepackage{comment}
\usepackage{dcolumn}
\usepackage{bm}
\usepackage{braket}
\usepackage{float}
\usepackage{multirow}
\usepackage{algorithm}
\usepackage{algpseudocode}
\usepackage{qcircuit}
\begin{document}


\title{Fault-tolerant Compass Codes}

 \author{Shilin Huang}
 \email{shilin.huang@duke.edu}
\affiliation{Department of Electrical and Computer Engineering, Duke University, Durham, NC 27708, USA}
\author{Kenneth R. Brown}
\email{ken.brown@duke.edu}
\affiliation{Department of Electrical and Computer Engineering, Duke University, Durham, NC 27708, USA}
 \affiliation{Department of Physics, Duke University, Durham, NC 27708, USA}
\affiliation{Department of Chemistry, Duke University, Durham, NC 27708, USA}
\begin{abstract}
We study a class of gauge fixings of the Bacon-Shor code at the circuit level, which includes a subfamily of generalized surface codes.  We show that for these codes, fault tolerance can be achieved by direct measurements of the stabilizers. By simulating our fault-tolerant scheme under biased noise, we show the possibility of optimizing the performance of the surface code by stretching the bulk stabilizer geometry. To decode the syndrome efficiently and accurately, we generalize the union-find decoder to biased noise models.  Our decoder obtains a $0.83\%$ threshold value for the surface code in quadratic time complexity. 
\end{abstract}

\pacs{Valid PACS appear here}
\maketitle

\section{Introduction}
Fault tolerance plays a central role in scalable and reliable quantum computation~\cite{shor1996fault,aharonov1999fault,kitaev1997quantum,knill1998resilient}.  
One leading candidate for fault-tolerant quantum computation is the surface code~\cite{kitaev2003fault,dennis2002topological}, which lies in the family of topological subspace codes~\cite{bombin2014structure}. It has an estimated fault-tolerant threshold value around $1\%$~\cite{fowler2012towards}, and only requires local interactions. These appealing properties open a promising path towards large-scale quantum computation~\cite{monroe2014large,fowler2012surface,hill2015surface,takeda2019toward}. However, implementing a universal set of logical operations on subspace codes is a challenging task~\cite{zeng2011transversality,eastin2009restrictions,bravyi2013classification}. 
Another candidate with lower overhead is the subsystem Bacon-Shor code~\cite{bacon2006operator,aliferis2008fault}. It is arguably the best for demonstrating fault tolerance in the near term~\cite{debroy2019logical} due to several practical advantages. For example, one can measure the non-local stabilizers fault-tolerantly by either $2$-local measurements~\cite{aliferis2007subsystem} or bare syndrome qubits~\cite{li2018direct}.
Also, asymmetric Bacon-Shor codes with particular size can have transversal multi-qubit controlled-$Z$ gates~\cite{yoder2017universal}. 
However, without code concatenation, the Bacon-Shor code fails to have a fault-tolerant threshold~\cite{aliferis2007subsystem,brooks2013fault}.

By gauge fixing the Bacon-Shor code, a large class of subspace or subsystem codes can be constructed, which are referred to as compass codes in Ref.~\cite{li20192d}. The flexibility of the code construction has several applications. For example, it provides a template for constructing topological subsystem codes~\cite{bombin2010topological,bravyi2012subsystem,bombin2015gauge}. From an architectural viewpoint, a subfamily of compass codes referred to as heavy hexagonal codes have been proposed to
minimize frequency collisions and crosstalk errors on superconducting qubits~\cite{chamberland2019topological}.
Moreover, it has been shown that at the phenomenological level, when the noise model is biased, one can choose asymmetric gauge fixings to improve the threshold~\cite{li20192d}. 
However, the behavior of compass codes at the circuit level was not studied in the context of a low-overhead fault-tolerant scheme. 

In this work, we first show that for a subclass of compass codes with a Calderbank-Steane-Shor (CSS) structure~\cite{calderbank1996good,steane1996multiple}, it suffices to use single syndrome qubits to extract error syndromes fault-tolerantly. Our protocol is a generalization of the bare-syndrome-qubit scheme for the surface code~\cite{dennis2002topological,wang2009threshold,tomita2014low} and for the Bacon-Shor code~\cite{li2018direct}. Using our bare-syndrome-qubit scheme, we study the circuit-level performance of a class of topological subspace compass codes, which can also be viewed as generalized surface codes, under biased noise on data qubits. We show that if the gate error rate is low, one can still benefit from asymmetric gauge fixing even if the gate does not preserve the bias. We expect that the performance can be further improved with the help of bias-preserving gates or fault-tolerant gadgets~\cite{aliferis2008fault, webster2015reducing, stephens2013high, puri2019bias}, and decoding algorithms for biased noise~\cite{tuckett2018ultrahigh,tuckett2019fault}.

The bottleneck in the time complexity of our numerical simulation is the decoding process. 
Instead of using minimum-weight perfect matching~\cite{edmonds1965paths,dennis2002topological}, the standard decoding algorithm for the surface code, we adopt the union-find decoder~\cite{delfosse2017almost} to accelerate our simulation. However, the linear-time union-find decoder only works on unweighted decoder graphs. For the purpose of studying asymmetric errors, a weighted decoder graph is needed to capture the asymmetry. With the cost of increasing the time complexity to quadratic, we generalize the union-find decoder graph on weighted decoder graphs. For the surface code, under standard depolarizing gate errors and measurement errors, we observe that use of a weighted graph will improve the threshold value from around $0.54\%$ to around $0.83\%$, approaching the $1\%$ threshold value obtained with minimum-weight perfect matching.

\section{Background}

\subsection{Compass Codes}
The Bacon-Shor code with distance $n$, denoted by $\mathcal{BS}_n$, is a subsystem code encoding a single logical qubit into an $n \times n$ square lattice of qubits~\cite{bacon2006operator,aliferis2007subsystem,shor1995scheme}. For convenience, we label the rows and columns of the lattice from $1$ to $n$, denote the qubit on row $i$ and column $j$ by $q_{i,j}$, and denote an operator $O$ acting on $q_{i,j}$ by $O_{i,j}$. The stabilizer group of $\mathcal{BS}_n$, denoted by $\mathcal{S}$, is generated by 
$\prod_{i} X_{i,j} X_{i,j+1}$ ($1\le j < n$) and $\prod_{j} Z_{i,j} Z_{i+1,j}$ ($1\le i < n$). The gauge group of $\mathcal{BS}_n$, denoted by $\mathcal{G}$, is generated by $2$-local operators $X_{i,j} X_{i,j+1}$ ($1\le i\le n$, $1\le j<n$) and $Z_{i,j} Z_{i+1,j}$ ($1\le i<n$, $1\le j\le n$).

\begin{figure}[t]
    \centering
    \includegraphics[width = 1.0\linewidth]{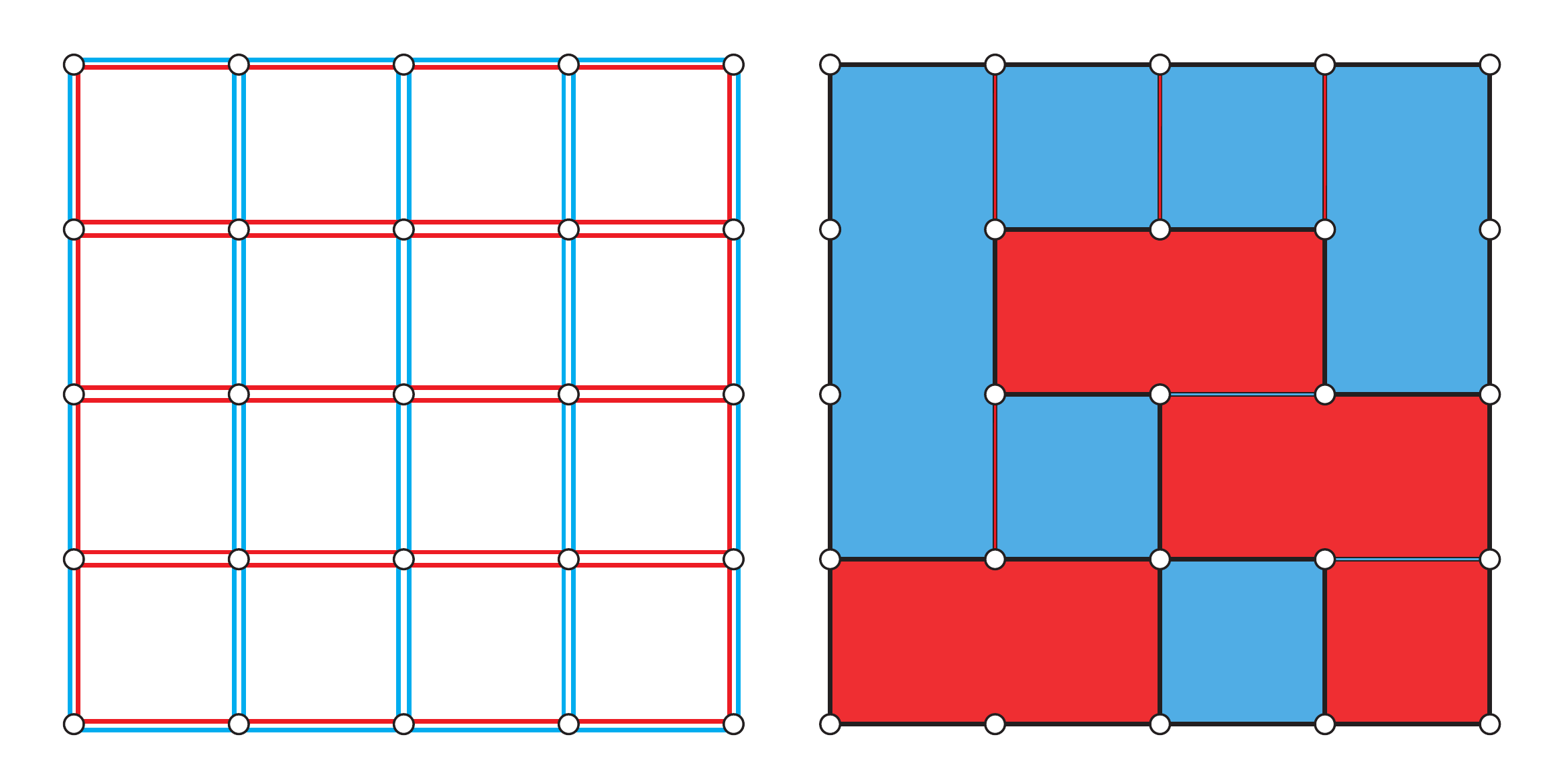}
    \caption{The construction of a $5\times 5$ compass code. We start with a $5\times 5$ Bacon-Shor code whose stabilizer group is generated by $\prod_{i} X_{i,j} X_{i,j+1}$ and $\prod_{j} Z_{i,j} Z_{i+1,j}$, which correspond to rectangles with blue and red boundaries, respectively. After gauge fixing the Bacon-Shor code, the stabilizer generators are cut by cells with the opposite color. For example, the stabilizer generator $\prod_{j} X_{j,2}X_{j,3}$ is cut by two red cells in the second column into three pieces $X_{1,2}X_{1,3}X_{2,2}X_{2,3}$, $X_{3,2}X_{3,3}X_{4,2}X_{4,3}$ and $X_{5,2}X_{5,3}$.}
    \label{fig:compass}
\end{figure}

Generally speaking, a compass code is a gauge fixing of the Bacon-Shor code~\cite{li20192d}. If the enlarged stabilizer group is maximal, we obtain a subspace code. One can also construct subsystem codes by leaving some gauges unfixed. We focus on a subclass of compass codes which can be easily visualized by coloring cells of the lattice. To describe this,
we first index each cell by the label of its top-left qubit; then we color each cell by red or blue, or leave it blank. 
If the $(i,j)$-th cell is red, we fix the gauge $\prod_{k=0}^i X_{k,j} X_{k,j+1} \in \mathcal{G}$. If it is blue, then we fix $\prod_{k=0}^j Z_{i,k}Z_{i+1,k} \in \mathcal{G}$. For those uncolored cells, there are no corresponding gauge fixes. Fig.~\ref{fig:compass} presents an example of a $5\times 5$ compass code. Note that since we only perform $X$- and $Z$-type gauge fixes, the resulting code is still a CSS code. Indeed, bit-flip ($X$-type) errors and phase-flip ($Z$-type) errors can be decoded separately. 
Importantly, an $X$-type ($Z$-type) error on a single qubit only changes no more than two syndrome checks.
This property guarantees that when we have perfect syndrome extraction gadgets,  decoding algorithms for the surface code~\cite{edmonds1965paths,fowler2013minimum,delfosse2017almost}  can be directly applied on compass codes. In fact, when all cells are colored, the resulting code is exactly a surface code defined on a planar graph~\cite{dennis2002topological}. In particular, the subspace Shor code and the rotated surface code correspond to a uniform coloring and a checkered coloring, respectively.

\subsection{Decoder Graph}\label{section:decoder_graph}
A standard approach for decoding the surface code is to build two decoder graphs to decode $X$- and $Z$-type errors separately. In the following we briefly review how to construct the decoder graph for $Z$-type errors only. The construction of the decoder graph for $X$-type errors is similar.

We first consider surface codes without boundaries. $X$-type syndrome checks are vertices of the graph, while qubits are edges. For each qubit, we link the two $X$-type syndrome checks with support on that qubit by an edge. $Z$-type errors form a collection of paths in the graph, and only the syndromes at endpoints of these paths will flip. The decoding problem then becomes finding the most probable collection of paths, given the endpoints of paths only. For symmetric noise models, an efficiently computable choice is the collection with the minimum total length, which can be further formulated as a minimum-weight perfect matching (MWPM) among the given endpoints~\cite{dennis2002topological,edmonds1965paths}. For each two endpoints, the weight of matching them is simply their distance in the decoding graph. For asymmetric noise,  one can weight each edge $e$ by $\log\left(\frac{1-p_e}{p_e}\right)$, where $p_e$ is the probability that a $Z$-type error occurs on the corresponding qubit~\cite{dennis2002topological}. Note that if two edges $e_1$, $e_2$ with error probabilities $p_1$, $p_2$ respectively link the same pair of vertices, then one can merge them as a single edge $e$, associating an error probability $p_e = p_1 (1-p_2) + (1-p_1) p_2$. 

For surface codes with boundary, there will be some qubits covered by only one $X$-type syndrome check, which leads to open edges in the decoder graph. Indeed, a path might have only one endpoint. To address this issue, one has to close those open edges by introducing extra vertices, which are allowed to be paired with the flipped syndromes. 

At the phenomenological level, i.e., when considering imperfect measurements, multiple rounds of syndrome extraction have to be applied, and the decoder should be able to capture measurement errors. To achieve this, we construct a new decoder graph with dimension $2+1$~\cite{dennis2002topological,wang2009threshold,fowler2012towards,tomita2014low}: for each round of extraction, we make a copy of the initial decoder graph to identify data errors in this round,
and for two adjacent rounds, we link the corresponding vertices together to represent measurement errors. 

\begin{figure*}[t]
\centering
\includegraphics[width = 1\linewidth]{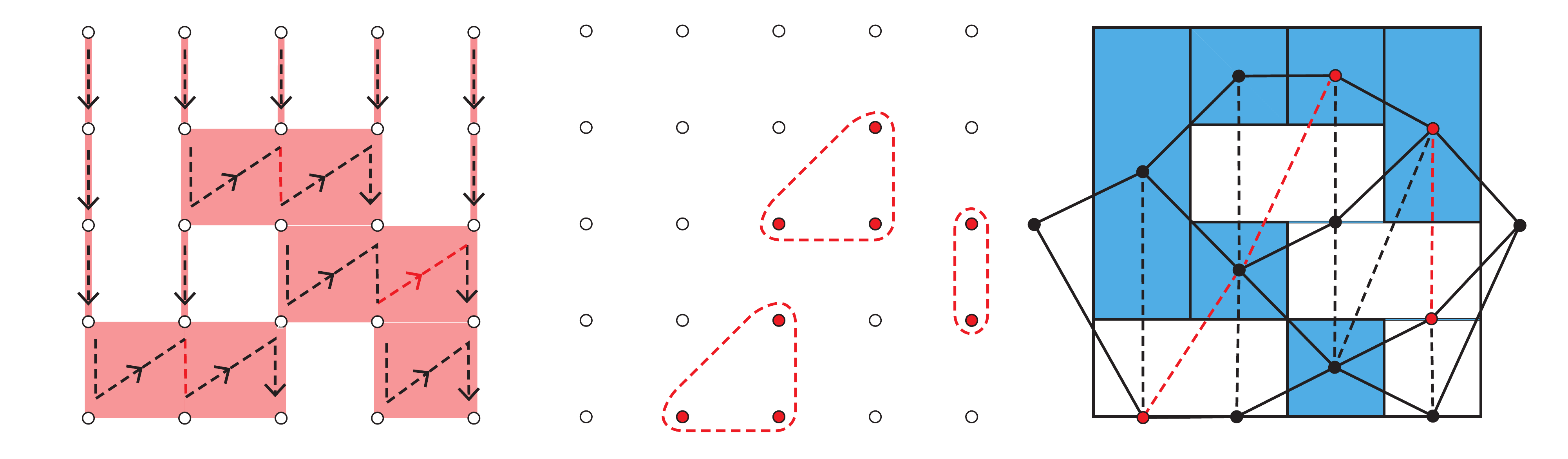}
\caption{Fault-tolerant syndrome extraction using bare-syndrome-qubits on a compass code.  On the left, each red rectangle represents a $Z$-type stabilizer.  CNOT gates are applied from data qubits to ancilla qubits in the order specified by the dashed arrows.  If a $Z$-type error occurs on the ancilla qubits after each CNOT gate represented by a dashed red line, those propagate to the three groups of data qubit errors circled in the middle. The right hand picture represents the decoder graph, with the three corresponding correlated errors highlighted in red.}
\label{fig:triangulation}
\end{figure*}
\section{Fault-tolerance with Bare Syndrome Qubits}
At the circuit level, the problem of decoding compass codes becomes more complicated: errors on the syndrome qubits might propagate to the data qubits through two-qubit gates and lead to high-weight data errors. Although universal fault-tolerant protocols can be directly applied so that syndrome-qubit errors become distinguishable~\cite{divincenzo1996fault,gottesman1998theory,steane2002fast,chao2018quantum,chao2018quantum,chamberland2018flag}, the resource overheads are usually prohibitive, and it will be difficult to represent these errors in the decoder graph. However, in the extreme case in which the stabilizers have the lowest weight, the surface code can be decoded fault-tolerantly with just bare syndrome qubits~\cite{dennis2002topological,tomita2014low}. As another extreme case, the Bacon-Shor code also has a bare-syndrome-qubit fault-tolerant scheme using a carefully designed gate sequence~\cite{li2018direct}. One naturally seeks to generalize such simple protocols on arbitrary compass codes, whose existence was shown in Ref.~\cite{li20192d}. 

To address this, we observe that in the language of compass codes, the fault-tolerant schemes for two different codes have a unified description:
for each $Z$-type stabilizer check $S = Z_{i,j_1}Z_{i+1,j_1}\cdots Z_{i,j_2}Z_{i+1,j_2}$ ($1\le i<n$, $1\le j_1\le j_2\le n$), we assign a syndrome qubit $a_S$ initialized to $\ket{0}$, and then apply controlled-NOT gates between data qubits and $a_S$, where the data are control qubits and $a_S$ is the target qubit.  Finally, we measure $a_S$ in the $Z$-basis. The ordering of the controlled-NOT gates has the following zig-zag pattern:

\begin{tikzcd}
q_{i,j_1} \arrow[d] &q_{i,j_1+1}\arrow[d]& \cdots\arrow[d] & q_{i,j_2}\arrow[d] \\
q_{i+1,j_1}\arrow[ur] & \cdots\arrow[ur,dashed] &  q_{i+1,j_2-1}\arrow[ur] & q_{i+1,j_2}
\end{tikzcd}\\
Although a single $Z$-type error on the syndrome qubit might propagate through the two-qubit gates, leading to errors of the form 
$Z_{i,j'}Z_{i+1,j'+1}\cdots Z_{i,j_2}Z_{i+1,j_2}$ or $Z_{i+1,j'+1}\cdots Z_{i,j_2}Z_{i+1,j_2}$, these errors will flip at most two checks that cover the two leftmost qubits, respectively. To represent these errors, one can add edges crossing the faces of the decoder graph. We notice that the new decoder graph is a triangulation of the original one. An important fact is that such a triangulation does not create any shortcut between two boundaries, which indicates that our scheme is distance preserving. See Fig.~\ref{fig:triangulation} for a demonstration of the bare-syndrome-qubit scheme working on the compass code in Fig.~\ref{fig:compass}.
Note that for the Bacon-Shor code, since its decoder graph is a chain without any inner faces, errors on syndrome qubits will not introduce any new edges. For the rotated surface code, all new edges are perpendicular to the logical-$Z$ operator. Therefore a shortest path between two boundaries will never cross these edges. Indeed, neglecting these edges in the decoder graph will not reduce the code distance. For the subspace Shor's code however, there exist shortest paths crossing the new edges, which demonstrates the necessity of the triangulation.

\begin{figure*}[t]
    \centering
    \begin{tabular}{ccc}
       \includegraphics[width=0.33\linewidth]{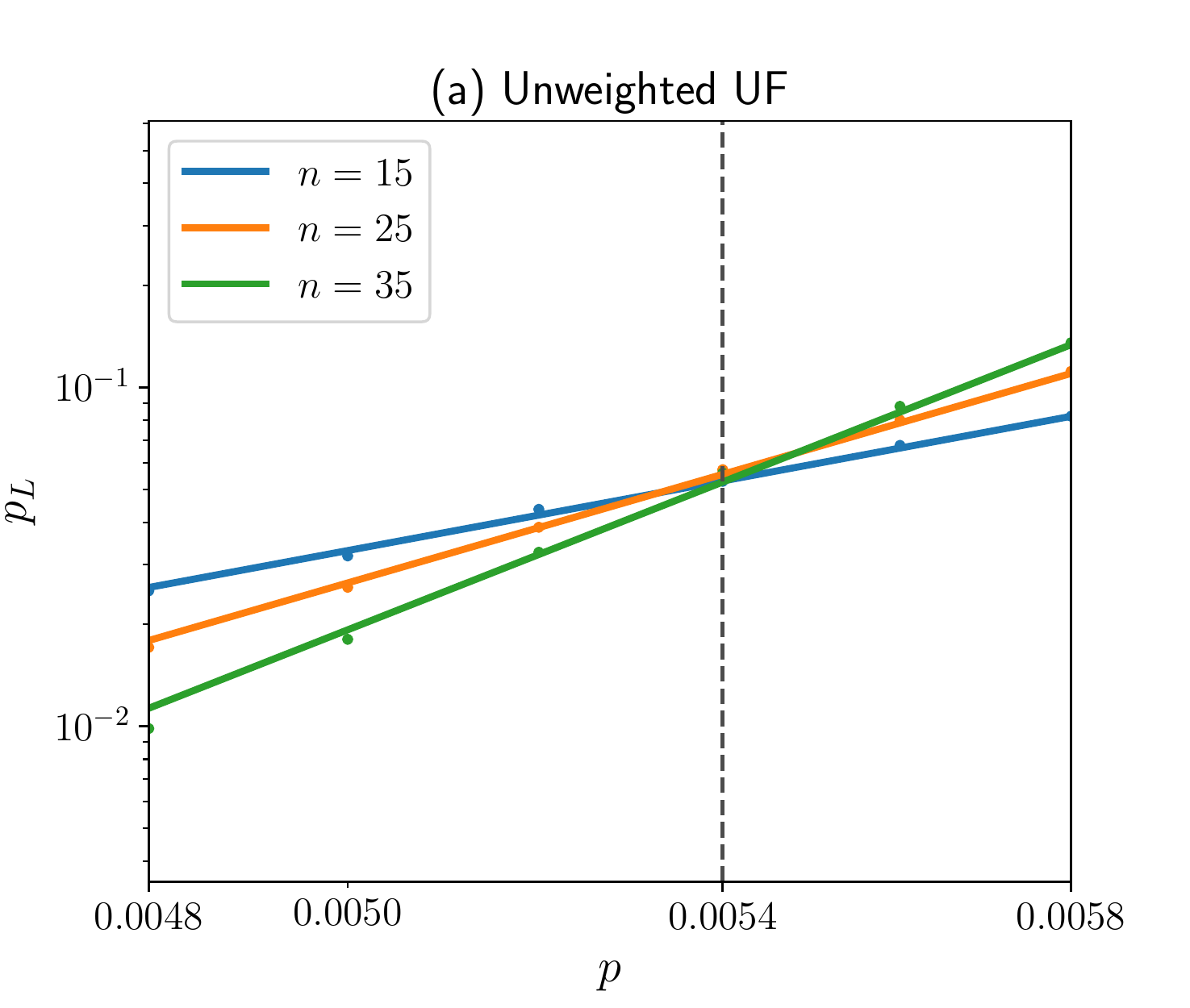} &
       \includegraphics[width=0.33\linewidth]{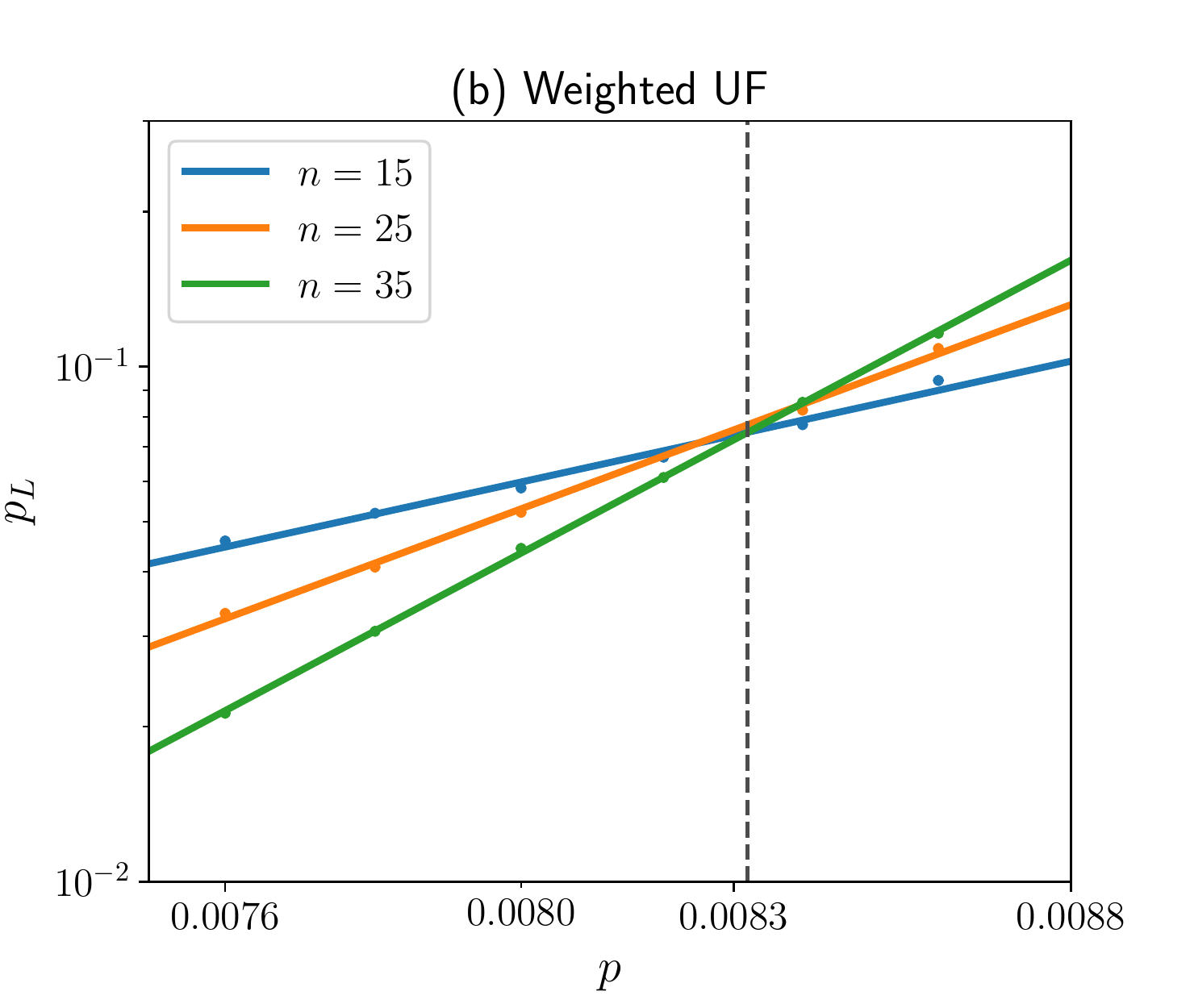}&
       \includegraphics[width=0.33\linewidth]{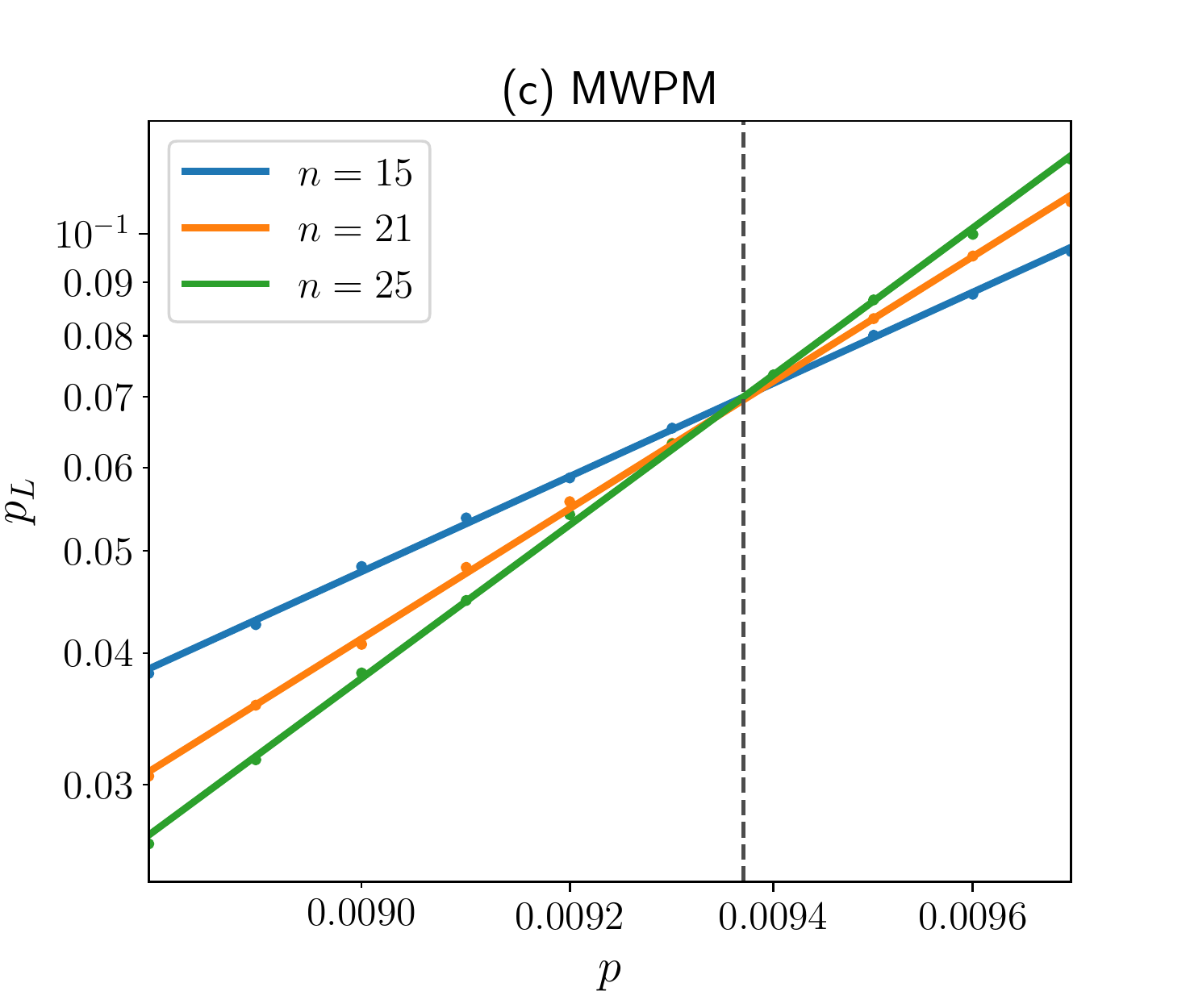}
    \end{tabular}
    
    \caption{Threshold behavior of the surface code with union-find decoders. Each two-qubit gate is followed by a two-qubit depolarizing channel with rate $p$, while each measurement fails with the same rate. The idling errors in the circuit are ignored. (a) If the decoder graph is unweighted, the threshold is around $0.54\%$. (b) Using a weighted decoder graph, we obtain a threshold value around $0.83\%$. (c) The threshold of minimum-weight perfect matching is around $0.94\%$ in this error model.}
    \label{fig:decoder}
\end{figure*}
\section{Decoder}

As mentioned in Section~\ref{section:decoder_graph}, our decoding problem can be formulated as a minimum-weight perfect matching (MWPM) problem, which can be solved by Edmonds' blossom algorithm~\cite{edmonds1965paths}. However, it will take $O(n^9)$ time to decode, since our decoder graph has $V = O(n^3)$ vertices and the time complexity of blossom algorithm is $O(V^3)$. To accelerate our simulation, we use the union-find decoder proposed by Delfosse and Nickerson~\cite{delfosse2017almost}. The idea of the union-find decoder is to greedily explore the metric space induced by the geodesic distance of the decoder graph. It finds a closed neighborhood of the flipped syndrome such that each path-connected component of that neighborhood either includes an even number of flipped syndromes, or intersects with the boundary of the surface. Finally, it pairs flipped syndromes by paths in the neighborhood arbitrarily. 
However, the union-find decoder only achieves almost-linear time complexity on unweighted decoder graphs. To decode the syndrome more accurately, we have to generalize it to weighted decoder graphs, at the cost of higher time complexity.

We first briefly describe the key part of the union-find decoding algorithm; that is, how to find a neighborhood which guarantees that the code distance can be preserved. The strategy is to use a greedy heuristic described by the following. Initially, the neighborhood is set to be the set of flipped syndromes only. In each step, we choose a path-connected component with the smallest boundary among those with an odd number of flipped syndromes. We then enlarge it by including the points whose distance from that component is no more than $\epsilon$. Here $\epsilon > 0$ is chosen to be the minimum value such that either the enlarging component intersect with another, or a new vertex is included. When two components intersect with each other, we merge them into a single component. 

Note that for unweighted graphs, components will only meet at either a vertex or the middle point of an edge. In this case one can always choose $\epsilon = 1/2$. Using the disjoint-set data structure for merging components~\cite{galler1964improved}, one can achieve $O(n^3\alpha(n^3))$ time complexity~\cite{delfosse2017almost}, where $\alpha(\cdot)$ is the inverse Ackermann function~\cite{tarjan1975efficiency}, which is less than $5$ for any practical value of $n$.
For weighted graphs, however, components can meet at any points in the metric space induced by the decoder graph. Without using advanced data structures, we have to determine $\epsilon$ by visiting all the edges on the boundary of the component, which takes $O(n^3)$ time. The total time complexity increases to $O(n^6)$, which is still significantly better than $O(n^9)$, the complexity of blossom algorithm. We estimate the threshold of the surface code with unweighted and weighted graphs under the following noise model: each two-qubit gate is followed by a two-qubit depolarizing channel with probability $p$, and each measurement fails with probability $p$. Note that we have not included preparation errors.
The simulation results are shown in Figure~\ref{fig:decoder}. We can see that decoding with weighted decoder graphs improves the threshold value from $0.54\%$ to $0.83\%$. As a comparison, we obtain an $0.94\%$ threshold value with the use of the mininum-weight perfect matching decoder.

\section{Biased Noise Model}
\begin{figure*}
\centering
\begin{tabular}{ccc}
   \includegraphics[width = 0.33\linewidth]{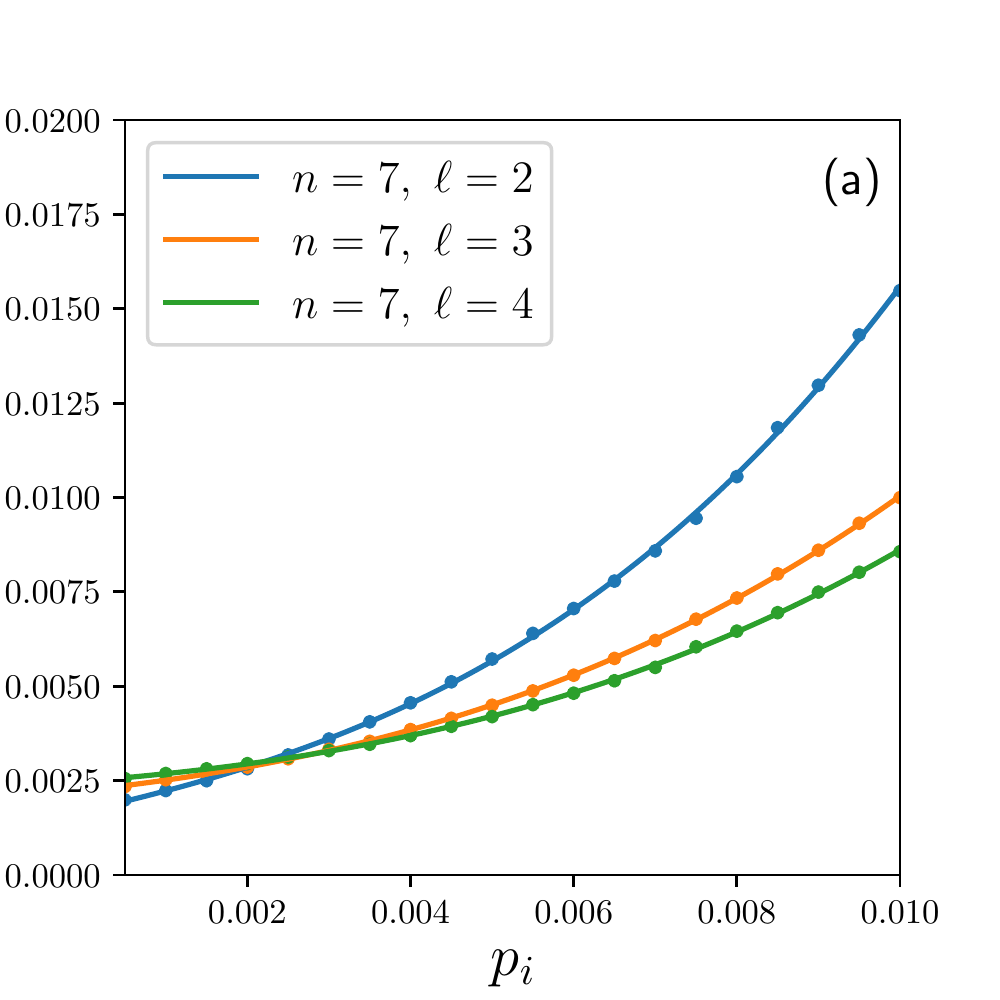} &
   \includegraphics[width = 0.33\linewidth]{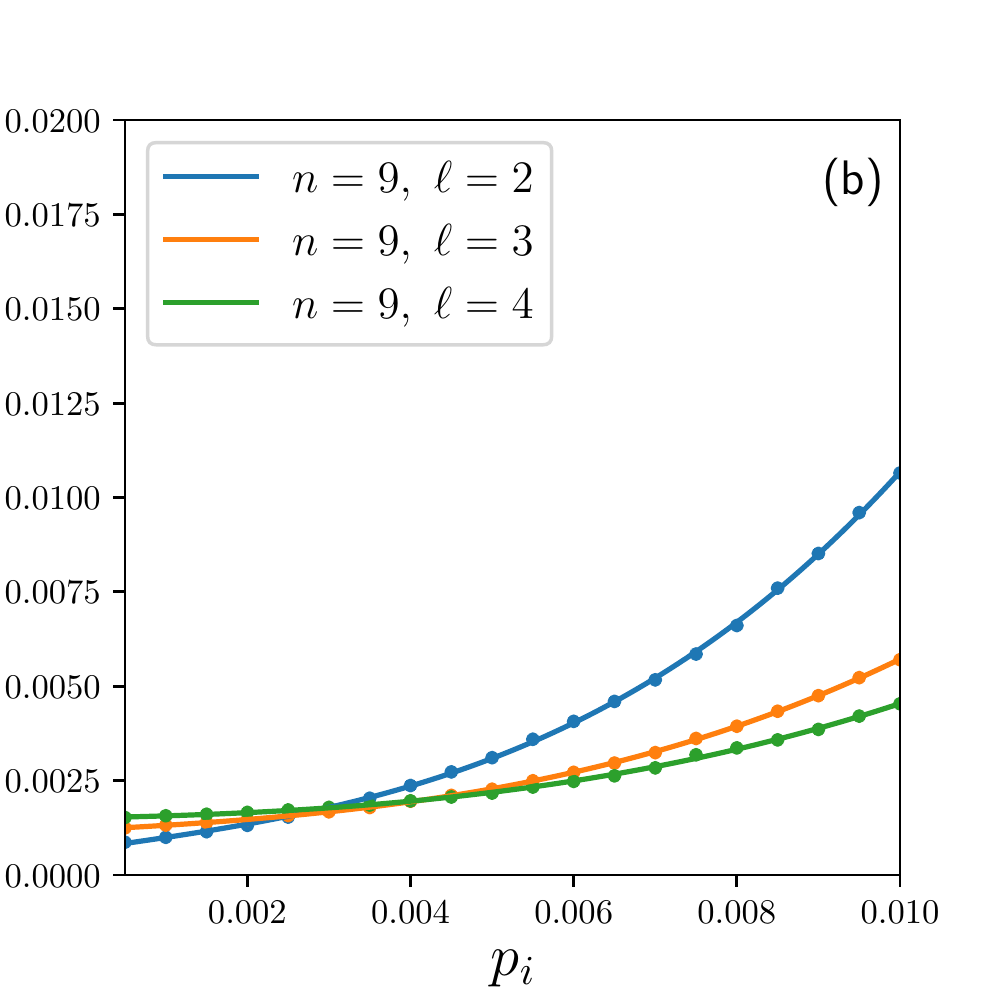} &
   \includegraphics[width = 0.33\linewidth]{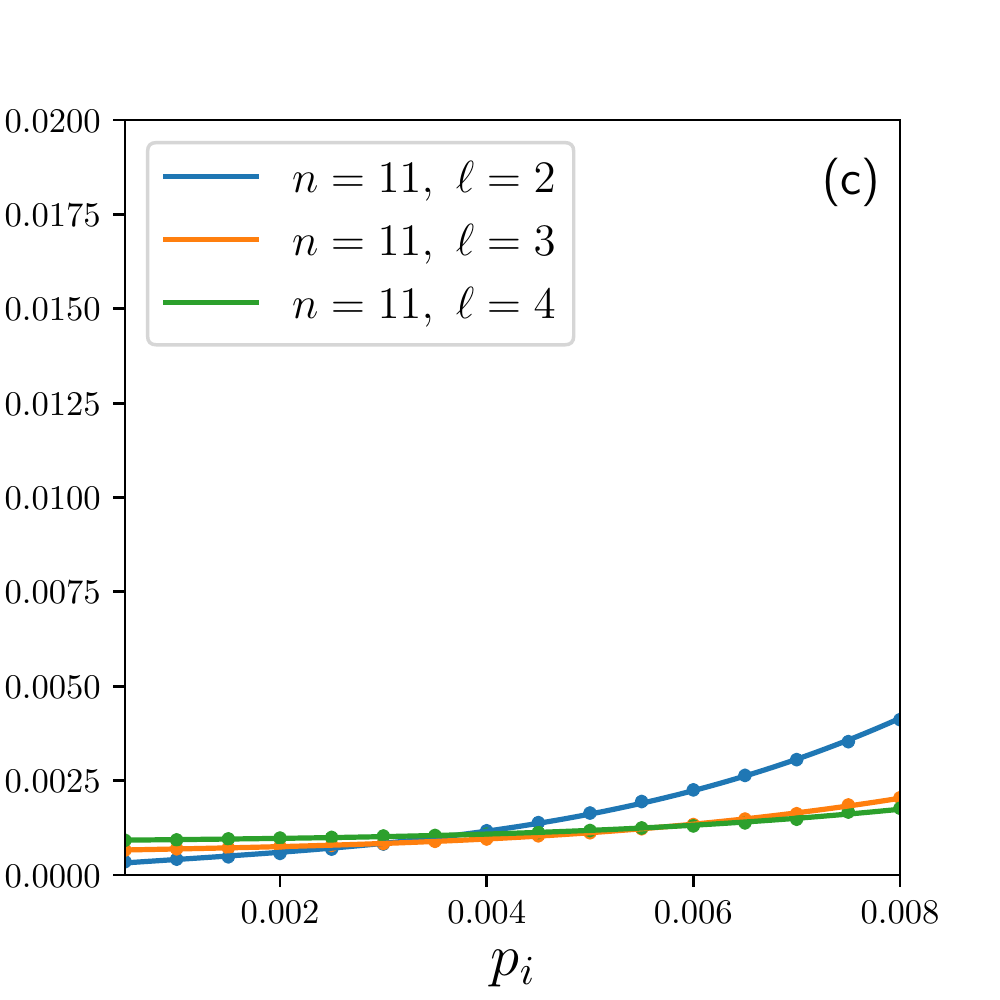}
\end{tabular}

\caption{Simulation results for intermediate-size elongated codes with fixed gate error rate $p_{gate}=0.3\%$ and measurement error rate $p_{meas} = 0.1\%$. $p_i$ is the idle error rate and $p_L$ is the corresponding logical error rate. (a) $7\times 7$ lattice. (b) $9\times 9$ lattice. (c) $11 \times 11$ lattice.}
\label{fig:result1}
\end{figure*}
In this work, we are interested in the performance of compass codes with pure dephasing memory errors and unbiased gate errors. The error model is as follows: each two-qubit gate is followed by a two-qubit depolarizing channel
\begin{displaymath}
\mathcal{E}(\rho) = (1-p_{gate}) \rho + \frac{p_{gate}}{15}\sum_{i} P_i \rho P_i,
\end{displaymath}
where $P_i$ runs through all non-trivial two-qubit Pauli operators; the outcome of each measurement flips with probability $p_{meas}$; each idling qubit experiences a dephasing channel
$\mathcal{E}_{idle}(\rho) = (1-p_i) \rho + p_i Z\rho Z$.
This follows from a $T_2$ dephasing model which commonly occurs in quantum devices~\cite{paik2011observation,wang2017single,groszkowski2018coherence}. If we fix $p_{gate}$ and $p_{meas}$, codes with a higher $p_i$ threshold will downgrade the requirement of long $T_2$ coherence time, or reduce the overhead by doing syndrome extraction less frequently.

If $p_{gate}$ and $p_{meas}$ are sufficiently small, the errors on data qubits will be biased to $Z$.
Previous work has shown that when the memory errors are $Z$-biased and the gate errors are ignored, one can improve the threshold scaling by fixing more $X$-type gauges~\cite{li20192d}. 
One might attempt to apply the same biased gauge fixing strategy at the circuit level. However, we note that errors on syndrome qubits will correlate excitations in different rows of the lattice, which makes the biased gauge fixing less effective. 

To study the effectiveness of biased gauge fixing at the circuit level, we focus on a subfamily of compass codes, which is referred to as elongated codes in Ref.~\cite{li20192d}. 
Here we recall that an $n\times n$ elongated code with elongation $\ell$ is constructed by coloring those cells $(i,j)$ with $i\equiv j\ (\mathrm{mod}\ \ell)$ by red, and the remaining cells by blue. As the elongation grows, more $X$-type gauges are fixed. 

We simulate the performance of $(2+1)$D elongated codes under our noise model. For a lattice of size $n$, we apply $n$ rounds of faulty syndrome extraction and an ideal round at the end, and then decode the syndrome with our decoder. We note that for codes with larger elongation, each syndrome qubit interacts with more data qubits.
To avoid the complication of scheduling, we assume that the dephasing errors only happen between two consecutive extraction rounds.

As a demonstration, we fix $p_{gate} = 0.3\%$ and $p_{meas} = 0.1\%$, and compare the performance among $\ell = 2,3,4$ for different values of the side length $n$ and the dephasing rate $p_i$. Note that $\ell = 2$ is the usual rotated surface code.  The simulation results for intermediate-size lattices are presented in Fig.~\ref{fig:result1}. 
We can observe that when the noise is unbiased, i.e., $p_i\rightarrow 0$, the surface code always performs better. For fixed $n$, when $p_i$ is greater than a critical value, the noise becomes sufficiently biased and elongated codes outperform the surface code. However, as the size of the lattice grows, the critical value increases towards the threshold of the surface code.

\section{Conclusion}
In this paper, we provide a simple fault-tolerant scheme for 2D compass codes with direct measurements, which is independent of the size of the stabilizer checks. Our scheme unifies the direct measurement schemes for the Bacon-Shor code~\cite{li2018direct} and the (rotated) surface code~\cite{tomita2014low}. With our protocol, we study the performance of compass codes under circuit-level noise. We show that for biased error models, it is possible to boost the performance with a biased gauge-fixing. We expect further improvements with the help of bias-preserving gates~\cite{puri2019bias}.

One drawback of our fault-tolerant scheme is the difficulty of parallelization of the circuit, which limits its practical use on systems with high idling error rate. However, it is always possible to reduce the circuit depth by applying cat state measurements~\cite{divincenzo1996fault}, and one can balance the circuit depth against the number of syndrome qubits. 

We also generalize the union-find decoder~\cite{delfosse2017almost} to weighted decoder graphs, which is of independent interest but also a crucial part for correcting biased noise on larger lattices. Our decoder can be considered as a greedy approach for minimum-weight perfect matching, which surely reduces the computational complexity. Note that the syndrome validation part of our algorithm has some similarity to Dijkstra's shortest path algorithm, which indicates the possibility of further reducing the time complexity by using advanced data structures~\cite{cho1998weight}. We leave this for future work.

\section*{Acknowledgements} The authors thank Michael Newman, Muyuan Li and Dripto Debroy for helpful discussions.  
This research was supported by ODNI/IARPA LogiQ Program (Grant No. W911NF-16-1-0082), ARO MURI (Grant No. W911NF-16-1-0349), and EPiQC - An NSF Expedition in Computing (Grant No. 1730104).
\bibliography{refs}

\end{document}